\documentclass[article]{revtex4-1}
\usepackage{xcolor}
\usepackage{graphicx}% Include figure files
\usepackage{dcolumn}% Align table columns on decimal point
\usepackage{bm}% bold math
\usepackage{comment}
\usepackage{hyperref}% add hypertext capabilities
\usepackage{tikz}
\usepackage{changepage}
\usepackage{braket}

\begin{document}

\preprint{APS/123-QED}

\title{
Understanding dense hydrogen at planetary conditions 
}% Force line breaks with \\

\author{Ravit Helled}
\affiliation{Institute for Computational Science, Center for Theoretical Astrophysics \& Cosmology, 
University of Zurich, Zurich, Switzerland}%
\author{Guglielmo Mazzola}
\affiliation{IBM Research Zurich, S\"aumerstrasse 4, 8803 R\"uschlikon, Switzerland
}%
\author{Ronald Redmer}
\affiliation{Institut f\"ur Physik, Universit\"at Rostock, A.-Einstein-Str. 23-24, 18051 Rostock, Germany}%
%\author{\RR{William J. Nellis}}
%\affiliation{Department of Physics, Harvard University, Cambridge, \RR{Massachusetts 02138}, USA}

\date{\today}% It is always \today, today,
             %  but any date may be explicitly specified

%%%%%%%%%%%%%%%%
\begin{abstract}
\bf{
Materials at high pressures and temperatures are of great interest for planetary science and astrophysics, warm dense matter physics, and inertial confinement fusion research. 
Planetary structure models rely on our understanding of the behaviour of elements (and their mixtures) at exotic conditions that do not exist on Earth, and at the same time planets serve as natural laboratories for studying materials at extreme conditions. 
In this review we discuss the connection between modelling planetary interiors and high-pressure physics of hydrogen and helium. First, we summarise key experiments for determining the equation of state and phase diagram of hydrogen and helium as well as state-of-the-art theoretical approaches. We next briefly review our current knowledge of the internal structures of the giant planets in the Solar System, Jupiter and Saturn, and the importance of high pressure physics to their characterisation. 
%Finally, we discuss future challenges and perspectives in both fields, which are currently flourishing, and their expected synergy. 
}
\end{abstract}

%\pacs{Valid PACS appear here}
\maketitle
%\tableofcontents

%  {\color{blue} 
\section{Introduction} 

Determining the compositions and internal structures of planets is a key objective in planetary science. 
% and efforts in this direction have been taken since several decades. 
Modelling planetary interiors, however, is not possible without knowledge of the behaviour of materials at high pressures and temperatures.  
The requirement to have a proper description of the EOS of various elements at planetary conditions sets the connection between the high-pressure physics and planetary science communities. 
%The link between planetary science and high pressure physics works in both ways: on the one hand, in order to construct internal structure models of planets, it is important to integrate existing experimental and theoretical information on the behaviour of materials at high pressures and temperatures. On the other hand, 
At the same time, planets are natural laboratories for material in exotic conditions, providing qualitative information about materials at high pressure. 
In this review, we concentrate on the gas giant planets Jupiter and Saturn that are mostly composed of hydrogen (H) and helium (He). 
These two elements account for about 85$\%$ and 75$\%$ of their total mass, respectively. 

H-He are the lightest and most common elements in the Universe, and they account for nearly all the nuclear matter. 
Nevertheless, their behaviour at high pressures is still not fully understood and is subject of intense research~\cite{mcmahon_properties_2012}.
Understanding the nature of the giant planets is linked to the equation of state (EOS) of H-He: for example, the bulk density of Jupiter and Saturn is larger than that of pure H-He composition. 
%if one draws the mass-radius of planets composed of hydrogen (H) and helium (He) and compares them to the radii of Jupiter and Saturn, 
From their densities it can already be concluded that both planets must include heavier elements (typically assumed to be rocks/ices), with Saturn being more enriched than Jupiter. The predicted total mass of heavier elements depends on the H-He EOS. This in turn leads to different inferred total heavy-element masses in the planets and their distributions (see section~IV for details).   

%As an example, let's consider Jupiter and Saturn: their density (computed from the observed radii and calculated masses) is consistent with pure Hydrogen-Helium objects. 
In addition, the fact that both Jupiter and Saturn have strong magnetic fields implies that the material in their deep interiors is electrically conductive. 
Since both planets are primarily composed of H it suggests that hydrogen, which is a non-conducting molecular gas at standard conditions, changes its behavior drastically when compressed to high densities and becomes a mono-atomic metal. Metallization of solid H has been predicted by Wigner \& Huntington~\cite{wigner} to occur at 25~GPa already in 1935 based on the nearly free electron model. The quest for metallic hydrogen has been a major stimulus for high-pressure research since then which has lead to many breakthroughs in high-pressure experimental technique, in particular using diamond anvil cells (DACs). Up to now five solid phases were identified with increasing pressure up to the 400~GPa region (for a recent review, see~\cite{Gregoryanz2020}): 
a hcp solid phase I which is a molecular insulator, a broken-symmetry phase II above 60~GPa and below 100~K with partially ordered molecules, and another hcp phase III with an unusually intense infrared activity. Further high-pressure solid phases were detected only recently. Phase IV has drastically changed optical properties and structure search studies suggest that it consists of alternating layers of six-atom rings and free-molecules. This phase transforms gradually into phase V between 275~GPa and 325~GPa at 300~K which is perhaps partially atomic, i.e.\ it is a precursor for mono-atomic, pure metallic hydrogen. The great influence of correlations in strongly compressed hydrogen, of quantum mechanical effects, and of the complex high-pressure structures as well as their mutual interconnections hinder a full understanding of the transformations from a molecular solid to a mono-atomic metal, in particular, the role of pressure-driven dissociation.  

Conducting fluid hydrogen was observed in reverberating shock-wave experiments 60 years later~\cite{weir1996metallization} which showed that magnetic field generation in Jupiter starts at about 140~GPa, i.e.\ much closer to its surface than previously thought.  
In this review, we focus on the gas giant planets Jupiter and Saturn and therefore the behavior of H (and He), but similar arguments hold for water and ammonia, simple compounds that we observe as insulating in our everyday life, but should be conductive at high pressures to explain the magnetic fields of Uranus and Neptune that are not H-He-dominated in terms of composition \cite{2011ApJ...726...15H}. 
\par

%\rh{Planetary gravitational and magnetic fields contain information about the natures of planets interiors, provided realistic planetary models can be calculated that match measured values and spatial geometries of those fields.}
%Moving forward from such qualitative considerations, 
The availability of accurate measurements of the planets' gravitational and magnetic fields requires the development of detailed structure models of the planets. The planetary internal structure is inferred from theoretical models that fit the available observational constraints by using theoretical EOSs for H, He, their mixtures, and the heavier elements. 
These models are guided by experiments and theoretical calculations of the thermodynamic properties of the relevant materials at high-pressure. 
Planetary interior models however, are non-unique and several of the assumptions made by the models rely on physical and chemical processes that are not fully understood (see section IV for discussion).
Nevertheless, modeling giant planets in not possible without having a reliable EOS for H.  
Since laboratory experiments cannot cover the entire phase diagram 
for all relevant compositions, especially at extreme conditions (see section~\ref{ss:experiments}), they need to be accompanied and complemented by theoretical calculations of the strongly interacting quantum systems of electrons and ions. % which is a challenge for many-particle theory too. 
Numerical simulations solve the corresponding N-particle Schr\"odinger equation using techniques like quantum Monte Carlo (QMC)  or Density Functional Theory (DFT)
 %\textit{ab initio} molecular dynamics (AIMD) 
simulations as discussed in section~\ref{ss:theory}.  
Figure~1 shows the hydrogen phase diagram  in the typical ranges of pressure and temperature relevant for the giant planets. 
%\rh{We indicate in the figure experiments performed at the phase boundaries. It should be noted, however, that }
%planetary science. For comparison we also plot the pressure-temperature (P-T) profiles of Jupiter and Saturn. 

This review provides an overview of the EOS of hydrogen, and hydrogen-helium EOS at planetary conditions, and the link between planetary interiors and high-pressure physics. 
In section II, we survey the progress in the experimental and theoretical/numerical fronts. In section III we discuss the behavior of hydrogen and a hydrogen-helium mixture at planetary conditions. 
Section IV briefly discusses how planetary interiors are models, focusing on Jupiter and Saturn. 
The current challenges and future are summarized in section V. 
%Section III We first survey XYZ, then discuss its relevance for ABC, before describing applications PQR'}.

% Uranus and Neptune, as well as qualitative phase diagram of H and other relevant materials.\todo{introduce briefly connection between physics of H and planetary science}

%%%%%%%%%%%%%%%%%%
\section{Hydrogen}
Compressed H is a system that has attracted the attention of both the theory and experimental communities for almost a century. 
%From a theoretical point of view, 
%H represents {\bf a simple} condensed matter system, and is therefore ideal for developing and testing quantum many-body models. 
Solid H has been predicted to metallize~\cite{wigner}, a fate shared with every other material under high enough compression, and to be a room-temperature superconductor~\cite{PhysRevLett.21.1748}, with a speculated phase transition into a super-fluid phase, driven by proton quantum effects~\cite{babaev2004superconductor}.
 % Progress in the compression techniques and theoretical methods devised in the quest to realize and characterize, respectively, metallic and superconducting hydrogen and hydrogen-rich materials at room temperature have been proven also very useful to study hydrogen at the much warmer planetary conditions.
 %\todo{introduce briefly connection between physics of H and planetary science}
Below we review the continuing experimental and theoretical efforts, focusing on breakthroughs of the last five years; earlier work is summarized in~\cite{mcmahon_properties_2012}. 
We then discuss the current consensus and open questions regarding the physics of dense liquid H.

%\section{Methods}
%\label{s:method}
\subsection{Experiments}
\label{ss:experiments}

%\rh{This section should be modified. there is a suggestion from the referee for the text...}\\
%\RR{I have shifted the text proposed by the referee which he took from a later paragraph in this section. 
%I have also adjusted the next paragraphs and tried to focus first on DACs and then on shocks. 
%And then mention the remaining stuff.} 
% RR: This more general paragraph can be skipped I think
%
%As mentioned earlier, static compression experiments, traditionally devoted to probe the room-temperature area of the H phase diagram, are now being used to characterise the warmer liquid phase.
%In static compression methods a sample is slowly compressed between diamond anvils~\cite{bassett2009diamond} (see~\cite{RevModPhys.90.015007} for review), and typically characterised (e.g., determine the crystal structure) through Raman and infrared spectroscopy~\cite{2003PhRvL..90q5701G,2000PhRvB..61.6535D,2015NatMa..14..495H,2019NatPh..15.1246E}, or x-ray scattering. 
%Metallization can be inferred from reflectivity~\cite{eremets2011conductive,Diaseaal1579} or direct electrical measurements~\cite{eremets2011conductive,drozdov2015conventional}.
%The Mbar pressure is also calculated from the diamond spectra\cite{eremets2003megabar,dalladay2016evidence}. 
Static compression setups using DACs allow for a precise control of the temperature, compared to their dynamic counterpart, since the  samples can be either cryogenically cooled~\cite{PhysRevLett.75.2514,Diaseaal1579}, and resistively~\cite{datchi2000extended} or laser heated~\cite{PhysRevLett.100.155701,dzyabura2013evidence,zaghoo2016evidence,zaghoo2017conductivity}.
While double-stage DACs can generally sustain record high pressures of $\sim 800$~GPa for metals~\cite{dubrovinsky2015most}, the high reactivity of dense H with the diamond limits the present achievable pressure at around $\sim 400$~GPa. Recently, Dias \& Silvera~\cite{Diaseaal1579} found evidence for solid metallic hydrogen in their experiments at 495~GPa.   
The highest conclusive pressure reached so far was 425 GPa \cite{loubeyre2020synchrotron}. 
Further discussions on the validity of this result can be found in 
~\cite{Goncharoveaam9736,Liueaan2286,commentloubeyre,eremets1702comments,geng2018public, Silveraeaan1215,2017arXiv170303064S}. 
Several groups have been recently achieved pressures exceeding 300 GPa at T below 300~K.
Such experiments revealed a very complex solid phase diagram featuring both temperature-driven~\cite{howie2015raman} and pressure-driven transition to novel solid non-metallic phases such as phase~V~\cite{dalladay2016evidence,eremets2016low} as well as phase $H_2$-PRE~\cite{dias2019quantum}.  

% To summarize, while static compression techniques lead to more controlled lab measurements, a precise characterization of H above 100~GPa remains experimentally challenging, \rh{and there is still on-going debate in the community on measuring H at these high pressures.} This is because reaching pressures beyond 100~GPa requires small samples, and since H itself is very light, there is a significant lack of diagnostic techniques that could be used to precisely characterise H.
%{\bf Drawing the phase boundaries.} 

% RR: I would follow the advice of Referee 4 for our summary instead of the above paragraph.
%\RR{Ravit, could you please insert the citations? I am not sure to what version they refer. Thanks.\\ 
To summarize, standard DACs have conclusively reached pressures of 425~GPa at room temperature and reveal at least five solid phases \cite{dalladay2016evidence,eremets2019semimetallic,loubeyre2020synchrotron,Gregoryanz2020} It remains inconclusive whether a metallic solid state has been reached \cite{Dias2017,Goncharoveaam9736,Liueaan2286,commentloubeyre,eremets1702comments,geng2018public, Silveraeaan1215,2017arXiv170303064S} with recent experiments suggesting rather a semimetallic state\cite{eremets2019semimetallic}. Due to issues relating to sample containment, high temperature studies in static experiments are challenging with few experimental diagnostics that can conclusively determine either melting or conductivity. 

Matter at extreme conditions 
%(e.g., high pressure $P$, density  $\rho = V^{-1}$ where $V$ is volume, internal energy $E$, temperature $T$ and entropy $S$ 
has also been generated with dynamic compression techniques using shock waves~\cite{2017umdc.book.....N,Nellis2019}. 
%since static compression experiments are traditionally limited to few megabar and lower temperatures.
%Concerning dynamical compression, while in the past such extreme conditions were achieved using intense X-ray sources generated near underground nuclear explosions, 
Modern experimental setups use, e.g.,  
(i) impact of flyer plates accelerated to velocities as large as 8 or 45~km/s generated with a two-stage light gas gun~\cite{weir1996metallization} or pulsed power~\cite{knudson2015direct}, respectively, (ii) high-energy optical lasers~\cite{Celliers677}, and (iii) spherical implosions driven by high explosives~\cite{Mochalov2018}.   

% RR: static-dynamic
Experiments on pure H and pure He were recently presented using a combination of both methods, i.e.\ DACs 
to pre-compress the sample and then laser-driven shocks to further compress the material.  
The inferred data for both materials were obtained at much higher pressure conditions than on the cryogenic sample \cite{2015JAP...118s5901B}. 

% RR: new DACs
Recent advances in static compression experiments using DACs have paved the way to reach even higher temperatures and pressures so that the thermodynamic properties accessible for the two techniques will overlap in the future, and therefore can be used for benchmarking. 

%{\bf Equation of state.} 
For decades dynamic-compression experiments were performed with sharp single-step shock waves.
The conservation laws of mass, momentum, and energy restrict the thermodynamic path that can be reached for given starting initial conditions of the sample to so called Hugoniot curve defined by 
\begin{equation}
\label{e:p}
   E(\rho,T) -E_0(\rho_0,T_0)=\frac12 \left( P(\rho,T) -P_0(\rho_0,T_0) \right) \left( \rho^{-1} -  \rho_0^{-1}  \right)~,
\end{equation}
%\begin{equation}
%P_H-P_0 = \rho_0(u_s - u_0)(u_p - u_0)
%\label{e:p}
%\end{equation}
%\begin{equation}
%V_H = V_0[1-(u_p - u_0)/(u_s - u_0)]
%\label{e:v}
%\end{equation}
%\begin{equation}
%E_H-E_0 = 0.5(P_H + P_0)(V_0 - P_H)
%\label{e:e}
%\end{equation},
where $E(\rho,T)$, $P(\rho,T)$, $\rho$, $T$, are internal energy, pressure, density, and temperature, respectively, and $0$ indicates the initial state.
%where mass density $\rho_H=1/V_H$, initial density $\rho_0=1/V_0$, and $u_0$ is initial particle velocity in front of the shock wave.  $H$ subscripts denote values on the Hugoniot.  $P_H$, $V_H$ and $E_H$ are generally determined by measurements of $u_s$ and $u_p$, or by measurement of $u_s$ and determination of $u_p$ by shock-impedance matching\cite{nellis_2017}. 
In dynamic compression experiments, 
$P$, $V$ (the volume) and $E$ are generally determined by measurements of the velocities of the moving plate and by shock-impedance matching~\cite{2017umdc.book.....N,Nellis2019}. 
This fact is of crucial significance because velocities are readily determined and thermodynamic states $P$, $V$, and $E$ cannot be determined directly at high pressures.
While the thermodynamic relation computed through Eq.~\ref{e:p} remains the sole source of absolute EOS data at pressures above $\sim$5~Mbar, 
dynamic compression experiments also face challenges.
%\RR{\textbf{We should be fair with the experiment and state what is feasible today:}} 
Usually, determining the density has an uncertainty linked to the interferometric velocity measurement. 
%\textbf{Typically few percent error -- Bill?}} 
%In addition, %and, equally importantly, 
A direct measurement of temperature is challenging, e.g., via streaked optical pyrometry (SOP)~\cite{SOP}, 
and is therefore often calculated from the spectrum of the emitted radiation, or estimated from theoretical models. 
\par

Given the limited range of pressure-temperature ($P$-$T$) that can be explored with this technique, the key of the Hugoniot data is to benchmark EOSs inferred from empirical models or first-principles simulations. 
%Concerning H, 
%this task is challenging because Hugoniot curves have been traditionally affected by large error bars, and systematic errors between the different setups allows only for a qualitative consensus regarding the compressibility of hydrogen and deuterium. To date, 
The most accurate dynamical compression experiment with D reported by Knudson \& Desjarlais~\cite{PhysRevLett.118.035501} using pulsed power is affected with an error of only $\sim2\%$ in density which allows to discriminate among  various theoretical models. 
A similar accuracy has been obtained earlier using the same technique for water~\cite{Knudson2012}. 
%Therefore, despite the experimental efforts, the data inferred from shock compression experiments  is not accurate enough to benchmark theoretical models. 
Nevertheless, other shock-compression techniques exhibit larger uncertainties so that a first-principles EOS for H valid for a wide range of $P$ and $T$ conditions as occurring in gas giants is rewarding. %\textbf{Mention already here wide-range EOSs, e.g., Sesame, SCvH, REOS etc?}}
%Indeed since current quantum Monte Carlo simulations\cite{PhysRevLett.115.045301}, the most accurate class of method at our disposal to calculate EOS (cfn. Sect..), are still displaying a disagreement of about $10\%$ compared to Ref. \cite{PhysRevLett.118.035501}, it's not possible to determine if errors are coming from the experimental side, the theoretical, or both.

%{\bf Probing the phase diagram.} 
In order to explore a larger portion of the thermodynamic space, and to reach higher densities while keeping the temperature relatively small, dynamical compression featuring multiple shocks or ramp compression techniques are now being used. %\RR{\textbf{Add references?}} 
%\rh{Lowest temperatures and highest densities achieved  along isentropes, that can be approximated by multiple-shock compression pulses. Quasi-isentropic compression of a liquid with low mass density, such as H$_2$, is achieved by a shock wave reverberating in the sample contained between two dense, strong, weakly-compressible  anvils.Each of the individual small jumps in pressure contributes some shock-dissipation energy, with the result of keeping the sample temperature sufficiently small.}
The most recent experimental setups to study metallization of liquid H~\cite{knudson2015direct,Celliers677}, and high-pressure phases of water-ice~\cite{millot2019nanosecond} employ shock reverberation techniques.

\indent % The experimental methods to characterise the H phase diagram vary with the phases and the compression techniques. 
The solid-liquid boundary can be assessed either via direct sample observation~\cite{PhysRevLett.100.155701},
%observing laser speckles, 
by tracing the plateau in the temperature versus laser heating curve~\cite{PhysRevLett.100.155701} (latent heat), or from the disappearance of the Raman-active lattice modes~\cite{eremets2009evidence,Subramanian6014,PhysRevLett.119.075302}. Currently, the validity of both methods stated has been questioned by other research groups and should be further explored. 
Nevertheless, all recent experiments agree on the presence of a maximum in the melting line at about 100~GPa, and the liquid phase is always found to be stable above $\sim1000$~K \cite{zha_high-pressure_2013,howie2015raman}. 
This finding rules out the existence of compressed solid molecular hydrogen in the deep interior of giant planets.  
However, liquid H at planetary conditions undergoes 
%other important transitions, that is 
molecular dissociation and metallization. 
Resistivity has been also directly measured using metal electrodes, in the pioneering work of Weir \textit{et al.}~\cite{weir1996metallization}. 
%However, the value of the electrical conductivity reported is 6-8 times smaller compared to the recent work of Zaghoo \textit{et al.}~\cite{zaghoo2017conductivity}. 
%\RR{\textbf{I think this experiment is misleading and should not be mentioned here, possibly removed from our bib.}}
%Since the change in the optical properties is smooth, is associated to a conductivity threshold above the minimum metallic value of $\sigma_{DC} \approx 2000$ S/cm, it is clear that such uncertainty can shift this estimate by about $\sim 100$ GPa. 
A large uncertainty in the measurement of the optical properties also leads to an uncertainty in the estimated metallisation pressure ($\sim$100~GPa), which is typically defined as the pressure at which the conductivity of the system reaches the minimum metallic value of $\sim$2000~S/cm as introduced by Mott~\cite{Mott1961} for $T=0$~K. \\
\indent Different experimental studies aimed at identifying these transitions by searching for possibly abrupt changes in several properties of the samples. This is, however, an indirect way the characterise the transition, whose mechanism has to be interpreted from the data.
Since discontinuous changes in density, fingerprint of a first-order phase transition, are not observable in a fixed-volume DAC device, a phase transition in the liquid has been claimed from heating plateaus in static compression experiment~\cite{dzyabura2013evidence,ohta2015phase,zaghoo2016evidence}. This information alone is insufficient to assess metallization. 
Electrical conductivity can be estimated from reflectivity and the absorption coefficient in both static~\cite{zaghoo2017conductivity} and dynamic compression experiments~\cite{Celliers677,knudson2015direct,zaghoo2017conductivity}.

%%%%%%%%%%%%
\subsection{Numerical Calculations}\label{ss:theory}
Nowadays computer simulations are essential to understand fundamental physical problems %natural phenomena 
and assist the experimental realization of novel materials. In the case of H at high pressure, computer studies also assist and complement experiments in understanding its rich phase diagram. 
%(especially in the range of pressures today inaccessible).
Unlike experiments, in simulations one can monitor the ionic structures produced, and therefore, it is fairly straightforward, %easy, 
within the approximation of choice, to compute the relevant structures, the corresponding EOS, and the phase diagram.
%It is also straightforward to explore different chemical compositions.

Before the availability of powerful computers, the planetary physics community relied on semi-empirical models, employing classical force-fields fitted against available experimental data~\cite{ross1983equation}. A very successful semi-empirical EOS for H and He that covers a large range of temperatures and pressures is Saumon-Chabrier-Van~Horn (SCVH)~\cite{saumon1995equation}, which qualitatively reproduces several liquid-phase experimental data from low- to high- pressure. This widely-used EOS in astrophysics which includes H, He and their mixture, has been recently updated~\cite{2019ApJ...872...51C} and is in good agreement with existing experimental data and numerical simulations.  
Other frequently used EOS are those of Ross~\cite{PhysRevB.58.669} and Kerley~\cite{kerley2003}.  
Modern EOS tables are inferred from \textit{ab initio} simulations using Density Functional Theory (DFT) in the warm dense matter regime %vicinity of the dissociation region 
(see below), rather than from experiments. The most recent versions are from the early 2010's, by Caillabet \textit{et al.}~\cite{PhysRevB.83.094101}, Militzer and collaborators~\cite{militzer2013ab,2013PhRvB..87a4202M} and Becker \textit{et al.}~\cite{becker2014ab}.

Indeed, the quantum mechanical treatment of the constituents, electrons, and protons is required. 
First-principles calculations are essential to model the interesting regime of pressures between 10 and 1000~GPa, and temperatures below $10^5$~K.
These temperatures are sufficiently low that electrons cannot be modeled as an uniformly distributed background, and the system cannot be described by a two-component plasma~\cite{brush1966monte}.
The dissociation and metallization transitions occur at pressures and temperatures of the order of 100~GPa and 1000~K, respectively, and the effects of electron correlations and their interactions in a non-uniform ionic lattice are particularly complex, and therefore very accurate electronic structure solvers are needed. Another source of complexity is linked to the protonic quantum effects that must be included at $\sim$1000-2000~K, and the protons cannot be treated as classical point-like particles~\cite{morales_nuclear_2013}.

The commonly-used method to simulate materials at $T$ below $10^4$-$10^5$~K is 
\emph{ab initio} molecular dynamics (AIMD), which is basically an iterative scheme~\cite{allen_computer_1987} involving the alternate solution of the Newton (for the nuclei, if treated classically) and the non-relativistic Schr\"odinger (for the electrons) equations.
A notable exception, relevant for hydrogen, is the Coupled Electron-Ion Monte Carlo (CEIMC) method~\cite{pierleoni_coupled_2004}, where ionic displacements do not follow the equation of motions, but are determined via Monte Carlo sampling.

{\it Ab initio molecular dynamics} is based on the Born-Oppenheimer approximation that decouples the ionic and electronic degrees of freedom~\cite{RevModPhys.64.1045,alavi2009ab}. 
The electronic structure is computed (approximately) at fixed ionic positions, then the nuclei are moved according to the forces, evaluated from the electronic ground state. This is a second approximation, called ground-state Born-Oppenheimer, which is valid in the range of temperatures considered. The accuracy of the forces depends on the degree of precision employed to solve the electronic problem.
Thermo-states or baro-states are applied on the ionic degrees of freedom to sample from the desired constant-temperature (NVT), or constant-pressure (NPT) ensemble (see, e.g.,~\cite{andersen1980molecular,allen_computer_1987,martyna1994constant,bussi2007canonical}).
Zero point motion of the nuclei, due to their quantum nature, can be included in a straightforward manner using path-integral methods~\cite{PhysRevB.36.2092,cao1994formulation,ceperley1995path,morales_nuclear_2013, pierleoni2016liquid}.
%Solving the ground state electronic problem almost exactly, using quantum chemistry methods, is possible only for very limited sizes, consisting of dozen of electrons. 
%Several tools now exist to tackle this problem on reasonably large system sizes (e.g. 100-1000 electrons), the first and most popular is
The foundation of the AIMD method began with the pioneering work of Car \& Parrinello~\cite{PhysRevLett.55.2471}
in 1985, where for the first time, a first-principles electronic structure calculation such as Density Functional Theory (DFT)~\cite{hohenberg_inhomogeneous_1964,kohn_self-consistent_1965} was combined with molecular dynamics.
 % , which can tackle systems having order of thousands valence electrons, but with less accuracy compared to quantum chemistry methods.
 %The second is quantum Monte Carlo (QMC), which recently  emerged as promising tool to solve the electronic problems as it provides a good balance between accuracy and size.
% The conceptual distinction between the two is that QMC aims to solve approximately the exact Schr\"odinger equations, whereas DFT can solve exactly an approximate one.
 
{\it Density Functional Theory} aims to solve % a mean-field equation which targets the 
the $N$-particle Schr\"odinger equation via the exact electronic density instead of the  many-body wavefunction. 
This method leads formally to the exact result, but the explicit functional introduced in the Kohn-Sham equations to describe exchange and correlation (XC) effects between electrons is unknown~\cite{kohn_self-consistent_1965}. This key problem of DFT is subject of intensive work which has led to successively improved approximations~\cite{burke_perspective_2012}.  
%but a systematic and efficient route to improve the XC functional is still lacking~\cite{burke_perspective_2012}. 
AIMD simulation results must be compared to available experimental data in order to benchmark their accuracy.  
Here we restrict the discussion to XC functionals that have been successfully used for H (see~\cite{cohen2011challenges,burke_perspective_2012} for details on the success and challenges of DFT). 
The Perdew-Burke-Ernzerhof (PBE)~\cite{pbe} XC functional has been chosen for the first simulations of the dense liquid, starting from the pioneering work of Scandolo~\cite{scandolo2003liquid} to more recent work~\cite{lorenzen,PhysRevB.75.024206,tamblyn2010structure,morales_evidence_2010}, and solid H~\cite{bonev2004quantum,pickard2007structure,liu2012room,PhysRevB.87.174110,PhysRevB.85.214114,naumov2015chemical,PhysRevLett.120.255701}. 
H-He mixtures have been studied with PBE as well~\cite{Morales03022009,PhysRevLett.102.115701}.
While PBE is still used, other and more sophisticated XC approximations have also been presented
%to rationalize experimental observations, more abundant in the solid phase: form the 
such as the Becke-Lee-Yang-Parr (BLYP) approximation~\cite{blyp} used in Refs.~\cite{PhysRevLett.122.135501,PhysRevB.94.134101}, the 
HSE functional~\cite{heyd2003hybrid}, that is an approximation affected by a smaller self-interaction error, in Refs.~\cite{PhysRevLett.122.135501,morales_nuclear_2013}, and functionals that include van-der-Waals long range interactions, such as vdw-DF1~\cite{vdwdf1} and vdw-DF2~\cite{vdwdf2}, in Refs.~\cite{morales_nuclear_2013,knudson2015direct,azadi2017role}.

The performance of first-principles simulations has been checked against the most accurate Hugoniot data observed so far at Sandia's Z machine~\cite{PhysRevLett.118.035501}.
The experimental data for the dissociation region are in disagreement with recent QMC calculations and better described by DFT simulations. 
However, none of the XC functionals is able to describe the complete experimental data set, but the vdW-DF1 functional seems to perform best. 
%\textbf{We should show the Hugoniot data from that paper.} 
This finding has been confirmed recently by Knudson \textit{et al.}~\cite{Knudson2018} in re-analyzing the pioneering multiple shock wave experiments 
of Weir \textit{et al.}~\cite{weir1996metallization}.
Figure 2 presents Hugoniot curves (Eq. 1) for H and obtained when using different theoretical EOSs and compared with the experimental data.

%Unfortunately, the inferred EOS strongly depends on used functional. 
%Systematic errors 
Predictions of different XC functionals differ by up to 100-200~GPa concerning phase boundaries in the solid~\cite{azadi_fate_2013}, 
and in the liquid~\cite{morales_nuclear_2013,knudson2015direct}.
Calculated reflectivities and conductivities vary by about a factor of three across this range of approximations, and the EOS by $\sim 10\%$~\cite{morales_nuclear_2013}.
Due to the lack of experimental data in the relevant $P$-$T$ range, the performance of XC functionals can only be benchmarked against first-principles theories such as QMC-based methods.
Recently a consensus has emerged that the vdw-DF1~\cite{vdwdf1} is the  best performing approximation in both the liquid~\cite{mazzola2018,pierleoni2016liquid}, and solid phases~\cite{PhysRevB.89.184106}, and is therefore the most commonly used DFT method in recent studies of H-He demixing~\cite{PhysRevLett.120.115703} and optical properties across the dissociation~\cite{rillo2019optical}.

{\it Electronic Quantum Monte Carlo} is a wave-function-based method~\cite{foulkes_quantum_2001} (unlike DFT), and is emerging as an accurate solver for electronic problems thanks to new generations of supercomputers. The main advantages of QMC compared to DFT are: (i) QMC relies on a many-body theory with a natural and explicit description of electron correlations, therefore the accuracy of calculations is in principle systematically improvable. (ii) QMC gives accurate results exhibiting, at the same time, a comparable scaling of computational cost with system size with DFT (although usually with a much larger prefactor). 
(iv) QMC is in an excellent position to take advantage of current and future supercomputer architectures, which are suitable for intrinsically parallel techniques.
%such as Monte Carlo. % rather than DFT\cite{Becca2017}.???? I am not sure it is true only for QMC...}

Nevertheless, QMC algorithms are still more computationally expensive than DFT, and as a result, they have been traditionally used for benchmarking purposes~\cite{PhysRevB.89.184106,clay2016benchmarking}, or to assess relative stability between a limited number of candidate solid  structures~\cite{chen2014room,PhysRevLett.112.165501,drummond2015quantum,azadi2018nuclear}.
The simulation of liquids is significantly more challenging since the QMC electronic solver needs to be coupled with an efficient sampling method for the ions. 
To date, two different strategies have been established: the first uses QMC in a molecular dynamics fashion 
(QMC-MD)~\cite{attaccalite_stable_2008,mazzola_unexpectedly_2014,mazzola_finite-temperature_2012,zen2015ab,mazzola2017,mazzola2018}, 
and the second, CEIMC, employs a multilevel sampling approach for both electrons and ions~\cite{pierleoni_coupled_2004,PhysRevLett.97.235702,morales_evidence_2010,morales2010eos,PhysRevLett.115.045301,pierleoni2016liquid}.

The QMC-MD approach is potentially indicated for the
simulation of large systems, as it does not rely on the computation of energy differences, that are extensive quantities, at the cost of introducing an integration time step error~\cite{attaccalite_stable_2008}. 
This approach has been employed for simulating other elements than H, such as water~\cite{luo2014ab,zen2015ab}, and H-He mixtures~\cite{mazzola2018}. The CEIMC method features a more sophisticated projective QMC solver, and can therefore achieve more accurate energetics. 
These two methods have been mainly used to simulate H dissociation and the metallization transition, and after some disagreements, they are now very consistent in predicting the nature and location of the dissociative transition in the liquid~\cite{pierleoni2016liquid,mazzola2018}. 
Hydrogen metallisation using QMC has been assessed either from direct conductivity calculations\cite{lin2009electrical} or from ground-state properties~\cite{mazzola2015distinct,pierleoni2016liquid}.

{\it Nuclear quantum effects} can be included in simulations using the path-integral formalism. 
In short, one point-like ion (with a classical representation) is replaced by a ring-polymer whose beads are interacting with suitable harmonic forces\cite{mcmahon_properties_2012}. The configuration of this extended space can be sampled with MD or with Monte Carlo methods, where the forces associated 
with the electronic structure can be also computed at the DFT level\cite{morales_nuclear_2013}. Including nuclear quantum effects is essential for predicting the solid phases\cite{mcmahon_properties_2012}, and for an accurate determination of the  first-order transition\cite{morales_evidence_2010, knudson2015direct, pierleoni2016liquid, zaghoo2018striking}. 
Table 1 lists the different computational methods and list their main strengths and weaknesses. 
%One major drawback of QMC simulations is that an efficient route to compute excited states is still lacking, which does not allow a direct calculation of some fundamental properties such as the conductivity, and the metallicity of the system must be inferred from an indirect ground-state calculation~\cite{mazzola2015distinct,pierleoni2016liquid}.

\begin{table*}[htb]
\centering
\def\arraystretch{1.4}
\scriptsize{
\begin{tabular}{ |c | c | c | c | } 
 \hline
 \hline
 & {\bf Chemical models} & {\bf Density Functional Theory} &

 {\bf Quantum Monte Carlo} \\ 
 \hline
{\bf Type} & semi-empirical & first-principles & first-principles \\ 
\hline
{\bf Target} & ionic force-fields & electronic density & electronic wavefunction \\ 
\hline
{\bf Strengths} & covers a large $P-T$ range   & a robust method   & a nearly exact method \\
  & providing thermodynamical properties &  large electrons number &  identifies well phase transitions \\
\hline
{\bf Weaknesses} & non first-principle,  not exact & not exact, results depend on XC functionals  & hard to benchmark, computationally expensive \\
  &  based on effective potentials   &  computationally expensive &   small electrons number \\
%&  & covers a relatively narrow  $P-T$ range & covers a narrow  $P-T$ range \\ 
%&  & thermodynamical  & covers a narrow  $P-T$ range \\ 
\hline
{\bf Simulation size} & --- & up to $5000$ electrons  & $\sim 100-200$ electrons \\
\hline
%{\bf Computational cost$^*$} & very low & can be performed on desktops & requires HPC facilities \\
\hline
\end{tabular}
}
\caption{The different computational methods for EOS calculation. }
%*for simulating a reference system contaning $100$ electrons with state-of-the art settings in the DFT and QMC case.}
\end{table*}

%%%%%%%%%%%%%%%%%%%%%%%%%%%%%%%%%%%%%%%%%%%
\section{H and H-He at planetary conditions}
\label{s:physics}
\subsection{Pure hydrogen}

%We review here the relevant experiments and simulations concerning the liquid phase of H, and the emerging picture based on these data.
At temperatures of a few thousand Kelvin, relevant for the deep interiors of Jupiter and Saturn, the dissociation and metallization transition of pure-H is a continuous process.
The most recent experiments performed at these conditions are reported by Davis \textit{et al.}~\cite{davis2016x}, where x-ray scattering measurements in dynamically compressed deuterium shows an onset of ionization at around 20-40~GPa and 3500-4500~K, and by McWilliams \textit{et al.}~\cite{PhysRevLett.116.255501} who found a continuous transition at these conditions. This picture is compatible with DFT~\cite{tamblyn2010structure,morales_evidence_2010} and QMC simulations~\cite{pierleoni2016liquid,mazzola2018}, which all report continuous transitions above 2000-3000~K. 
Most of the recent experiments and simulations provide evidences for of a first-order transition up to 1200-1500 K~\cite{zaghoo2016evidence,knudson2015direct, ohta2015phase, morales_evidence_2010,lorenzen, pierleoni2016liquid,mazzola2018}.
Nevertheless, a quantitative agreement on the location of the phase boundary, 
as well as a precise (within $1\%$ accuracy) determination of the EOS for pure-H, is still missing. 
Traditionally, the principal single-shock Huguniot has been identified as the benchmark for simulations. Interestingly, even when considering the experimental uncertainty discussed in Sect.~\ref{ss:experiments}, EOSs obtained with DFT are in better agreement with experiments compared to the QMC data, computed using CEIMC~\cite{PhysRevLett.115.045301}.
Recently, Clay \textit{et al.}~\cite{PhysRevB.100.075103} suggested that this behavior can be explained by a more favourable error cancellation in the variables $P$, $T$, entering the Rankine-Huguniot equation.
%Therefore it is not necessarily implied that DFT-based EOSs are more accurate than QMC ones.

We therefore suggest that the available EOSs for hydrogen should be benchmarked with the $P$-$T$-$\rho$ relations computed with the most recent QMC calculations by Pierleoni \textit{et al.}~\cite{pierleoni2016liquid} (CEIMC) and Mazzola \textit{et al.}~\cite{mazzola2018} (QMC-MD).
While a QMC-EOS table for a wide $P$-$T$ regime is still missing, the $P$-$\rho$ relation at $T=6000$~K computed in Ref.~\cite{mazzola2018} reveals a denser EOS by about $5\%$, compared to the current DFT-PBE based ones~\cite{militzer2013ab,becker2014ab}.

State-of-the-art \textit{ab-initio} setups feature
 simulation cells consisting of up to 512 (128) particles at the DFT (QMC) level of theory for production runs~\cite{lorenzen,Geng2019}. Electronic finite size effects errors are mitigated by  Brillouine zone sampling~\cite{PhysRevLett.102.115701,holzmann2016theory,Geng2019}.
This setup can provide satisfactory accuracy in the liquid phase. 
However, this can change in the solid phase and generally near phase transitions, where insufficient equilibration times or too small sizes can artificially stabilise one competing phase against another, e.g., delaying the onset of metallization, or hindering the possibility to distinguish a first-order from a continuous transition. 
In hydrogen this issue is amplified because the melting line and the dissociation transition between the molecular and the atomic liquid lie quite close in the phase diagram. 
%It has been recently shown that it can be really hard to distinguish between a liquid phase and an highly defective solid structure using AIMD particle 128 simulation in NVT ensamble, so that the liquid-solid transition can be easily confused with a liquid-liquid one\cite{cheng2019evidence}.

%\RR{\textbf{I disagree with the statement on large size simulations within AIMD. While Winfried considered already up to 2000 atoms in 2011, a recent paper considered about 3500 atoms~\cite{Geng2019}. Computing capacity increases, and new methods such as stochastic DFT will easily overcome limitations with respect to $N$. I have rephrased to:}}
% The inability of simulating very large sizes with AIMD, e.g., order of $10^4$ particles represents also a limitation when studying mixtures (e.g.\ miscibility of water~\cite{soubiran2015miscibility}, iron~\cite{soubiran2015miscibility}, and MgO~\cite{PhysRevLett.108.111101}, or decomposition of hydrocarbons~\cite{ancilotto1997dissociation} in hydrogen), especially when we need to establish possible phase separation of low stoichiometry heavy elements (e.g.\ mixtures to simulate the composition of the ice giant planets Uranus and Neptune~\cite{chau2011chemical}). Machine-learning based investigation are expected to bridge this gap in future simulations~\cite{G2019}. 
Simulating very large ensembles $\sim10^4$ particles with AIMD is not yet possible so that studies on mixtures is limited (e.g.\ miscibility of water~\cite{soubiran2015miscibility}, iron~\cite{soubiran2015miscibility}, and MgO~\cite{PhysRevLett.108.111101}, or decomposition of hydrocarbons~\cite{ancilotto1997dissociation} in hydrogen), especially when we need to establish possible phase separation of low stoichiometry heavy elements (e.g.\ mixtures to simulate the composition of the ice giant planets Uranus and Neptune~\cite{chau2011chemical}). Novel methods such as stochastic DFT~\cite{Cytter2019} or machine-learning based investigation are expected to bridge this gap in future simulations~\cite{G2019}.

%%%%%%%%%%%%%%%%%%%%%%%%%%%%%%%%%%%%%%%
\subsection{Hydrogen and helium mixture}
%The behaviour of pure-He at planetary conditions  is better understood than that of pure-H simply because  ionization of He requires much higher pressures and a phase transition is not expected to occur. \RR{\textbf{This is incorrect. Predictions for a PPT in He based on chemical models were made in the 1980ies by Ebeling et al. The MIT in He is currently discussed based on AIMD. Include this discussion here?}}
Despite its importance, there are very scarce experimental data concerning the phase diagram of H-He mixtures, which are still limited to room temperature~\cite{PhysRevB.36.3723,loubeyre1991new,PhysRevLett.120.165301,2018PhRvL.121s5702T}. 
%{\bf - Ravit: Hydrogen and Helium are the most common elements in the Universe, and their ratio...  For planetary modeling, the ratio between the two elements is set to be proto-solar (since the sun undergoes  hydrogen burning)}
H-He mixtures have been predicted to undergo a (temperature-driven) transition from a high-T fully miscible liquid, to a low-T phase-separated with helium-rich droplets~\cite{1977ApJS...35..239S,1977ApJS...35..221S,mcmahon_properties_2012}.
Therefore, the interaction between H and He under the interior conditions of Jupiter and Saturn leads to challenges in determining the EOS of mixture due to the expected immiscibility of He in H, results in He settling ("helium rain").  
This process, is expected to impact the planetary structure as we discuss in more detail in Sect.~\ref{s:models}.

The existing knowledge on the liquid phase and H-He demixing is based on simulations.
Most recent studies employ DFT at the PBE level, computing the free-energy under the ideal-mixing approximation for the entropy of mixing~\cite{PhysRevLett.102.115701}, and under the non-ideal one~\cite{morales2009phase}.
The latter reference is obtained with thermodynamic integration and is expected to be more accurate, with a shift (toward low-T) of the demixing temperature of order of 1000~K.
This computation was recently revisited using the more accurate vdw-DF functional and non-ideal entropy of mixing by Sch\"ottler \& Redmer~\cite{PhysRevLett.120.115703} who found lower demixing temperatures, slightly below the Jupiter adiabat (black solid) and crossing that of Saturn (black dashed) between 1.5 and 4~Mbar as shown in Figure~3. 
A change of optical properties upon demixing of H and He was predicted by Soubrian \textit{et al.}~\cite{2013PhRvB..87p5114S} and is expected to be tested by experiments in the near future. 
%Corresponding evolutionary tracks are currently calculated based on this new H-He phase diagram. \textbf{Cite Mankovich \& Fortney?}}
%Figure~2 shows the phase diagram for the H-He mixture~\cite{PhysRevLett.120.115703} as well as the isentropes of Jupiter (solid) and Saturn (dashed), respectively. 
%\todo{TODO: implications for giant planets, and possibility of no demixing in Jupiter.}

Phase separation is not the only important physical process that distinguish H-He mixtures from pure hydrogen. 
The presence of helium stabilizes the hydrogen molecules, delaying the onset of metallization towards higher densities. Such a shift has been calculated with QMC by Mazzola \textit{et  al.}~\cite{mazzola2018}: it amounts $\sim 70$~GPa at 1500~K and $\sim 30$~GPa at 6000~K. 
Therefore, the continuous transition from molecular to metallic H-He mixture in Jupiter's conditions is expected to occur between 40 and 60~GPa \cite{mazzola2018}.

%%%%%%%%%%%%%%%%%%%%%%%%%%%%%%%%%%%%%
\section{Modeling Planetary Interiors}
\label{s:models}
Traditionally Jupiter and Saturn were viewed as laboratories for understanding the EOS, however, as our knowledge has grown the importance of other complexities have arisen as we discuss below. 
Planetary models aim to determine the internal structure of planets, i.e., the composition and its depth dependence. %\RR{Profiles?}  
This is done by constructing interior models assuming various different structures and compositions that fit and reproduce the planetary observed properties such as the mass, radius, luminosity, and gravitational field data. 

Jupiter and Saturn are often referred as "gas giant planets", however, the bulk of their interiors are in fact not in the gaseous phase due to the high pressures and temperatures as can be seen in Figure~1. 
As a result, Jupiter and Saturn are in fact "fluid giant planets". While in their low-P atmospheres H is in a molecular form, and the densities are low, the deep interiors of the planets are characterised by condensed (liquid) H, and therefore 
one should think of these planets as "fluid planets".  
It should also be noted that gas giant planets do not have a surface like in the case of terrestrial planets,  and therefore there is no real transition between the planetary atmosphere and the deep interior. 
At the same time, there is no exact location where the atmosphere ends since the outer regions of the atmosphere are characterised by a very low-density gas. 
As a result, planetary scientists define the "surface" of a gas giant planet at the location where the pressure is 1~bar, similar to the pressure on Earth's surface. The temperature at this location is measured (with a small uncertainty). Then, with information of the temperature at 1~bar, the entropy of the outer envelope is determined and (adiabatic) models can be constructed~\cite{guillot_interiors_2005}. 
The basic observed properties of Jupiter and Saturn are listed in Table~2. Recent review chapters about the giant planets can be found in 
Refs.~\cite{FJ2016,MB2016,HG2018,RH2018} and references given therein.  
%Uranus and Neptune.  

The modelled planetary interior can consist of different layers, where the number of layers is often chosen to be the minimum number of layers (typically three) that is required to fit the observational data, and be consistent with our knowledge of the material properties. 
%Of course, this does not necessarily reflect the reality, and the small number of layered that is assumed is an outcome the relatively small number of observed parameters since there is no point in constructing a structure model with has many more free parameters than observed properties. 
The main sources of uncertainty in interior models arise from the assumed number of layers, the assumed composition and its distribution, the heat transport mechanism, and the dynamical effects that are typically neglected or treated in a simplified manner. Another important source of uncertainty is the physics of dense hydrogen (and helium) and the understanding of the nature of the phase transitions in that determine the planetary density profile. 

Planetary structure models are constructed assuming the planet is in hydrostatic equilibrium, and rely on conservation of mass and of momentum: 
%Planetary interior models are based on the assumption of mass and momentum conservation. 
\begin{equation}
\frac{dm}{dr} = 4\pi r^2\rho(r),
\end{equation}
\begin{equation}
\frac{1}{\rho(r)}\frac{dP}{dr}=-{Gm(r)\over r^2} + \frac{2}{3}\omega^2 r,% + {G M\over 4\pi \ R^3 r}\varphi_\omega,
\end{equation}
%\begin{equation}
%{dT\over dr}=\frac{T}{P}\frac{dP}{dr}\nabla_T,
%\end{equation}
%\begin{equation}
%{\partial u\over\partial t}+p{\partial\over\partial t}{1\over\rho}
%=q-{\partial L\over\partial m},
%\end{equation}
%this set of equations include the mass conservation, hydrostatic balance, and thermodynamic equations. 
where $m(r)$ is the mass within a sphere of a radius $r$, $P$ is the pressure, $\rho(r)$ is the density at $r$, $\omega$ is the planetary rotation rate (typically assumed to be constant, i.e., uniform rotation) and $G$ is the gravitational constant. 
In order to infer the temperature profile within the planet $T(r)$ the  thermodynamic equation ${dT\over dr}=\frac{T}{P}\frac{dP}{dr}\nabla_T$ must be solved. 
The determination of the temperature profile, however, is not simple, and depends on  the heat transport mechanism within the planet. This can be radiation, conduction, and convection, where the latter can also have more sophisticated  versions such as layered-convection. $\nabla_T$ relies on complex physics that is still being investigated \citep{LC12,LC13,DC19,Vazan16,Vazan18}.  
%and as a result, many structure models adapt the simple solution of assuming an adiabatic temperature gradient that corresponds to... {\bf bla bla...} 

For simplicity,  for a few decades Jupiter and Saturn were assumed to be fully convective, so the temperature gradient was assumed to be the adiabatic gradient $\nabla_{ad}$. % associated with convection. 
%ypically, the temperature gradient is taken to be the smallest among the adiabatic $\nabla_{ad}$, radiative/conductive $\nabla_{rad/cond}$ 
%%%This is because the temperature gradient is chosen to be the one for which the heat transport mechanism resulting in the smallest temperature gradient is the most efficient one. 
%The adiabatic gradient is given by $\nabla_{ad}$=$\frac{\partial lnT}{\partial lnP}\arrowvert_s$, where $S$ is the entropy, corresponds to a case in which the material is homogenous and convective. 
%The radiative/conductive gradient is given by $\nabla_{rad/cond} = \frac{3 \kappa L P }{64 \pi \sigma T^4 G m}$, where $\kappa$ is the Rosseland opacity which accounts for contributions from both radiation and conduction, and $\sigma$ is the Stephan-Boltzmann constant (see \cite{MB2016} and references therein for further details).
%The planet's density profile is set to reproduce the measured gravitational field of the planet, or more precisely, to reproduce the measured gravitational moments J$_{2n}$. 
% that are typically determined by spacecraft. 
%In addition, theoretical modellers assumed a relatively simple structure for the planets.
%/conduction/radiation. 
Last but not least, in order to model the planetary internal structure knowledge of the density dependency on the temperature and pressure, i.e., $\rho(P,T)$ is needed. This is essentially the knowledge of the EOS of the assumed materials.
Since both Jupiter and Saturn consist of mostly H and He, modelling their internal structures requires knowledge of the behaviour of these materials for a wide range of pressures (1~bar up to few ten Mbar) and temperatures (165~K up to a few 10,000~K).  
  
It should be mentioned that there is an alternative approach to model planetary interiors by constructing "empirical models"~\cite{1995JGR...10023349M,2000P&SS...48..143P,2009Icar..199..368H,2011ApJ...726...15H}. These models do not rely on a physical EOS, and instead, the density profile has a mathematical representation such as polynomials or polytropes. Then by assuming hydrostatic equilibrium the planetary $P$-$\rho$ profile that reproduces the planetary observed parameters can be found. 
%represented by a more mathematical 
Although in such models the temperature profile cannot be determined, the composition can be inferred from the $P$-$T$ relation combined with an assumed temperature gradient. 
The main advantage of this approach is that it is less biased by the modeller's assumption, and it can lead to a larger parameter-space of solutions. 
% and the solutions are less dependent on the assumed model parameters. 
%that prwould be consistent with the observed parameters. 
%{\bf more information}. The inferred pressure-density profiles can then be interpreted by comparing with.... 
\begin{table}[h!]
\def\arraystretch{1.}
\centering
\footnotesize{
\label{tab:1}   
\begin{tabular}{p{3.5cm}p{2.25cm}p{2.25cm}}
\hline\noalign{\smallskip}
 {\bf Physical Property} & {\bf Jupiter} & {\bf Saturn}  \\
 \hline
%\noalign{\smallskip}\svhline\noalign{\smallskip}
Distance to Sun (AU) & 5.204 & 9.582\\
Mass (10$^{24}$ kg)& 1898.13$\pm$0.19 &  568.319$\pm$0.057\\
Mean Radius (km) & 69911$\pm$6& 58232$\pm$6 \\
Equatorial Radius (km) & 71492$\pm$4& 60268$\pm$4 \\
Mean Density (g/cm$^{3}$) & 1.3262$\pm$0.0004 & 0.6871$\pm$0.0002 \\
$J_2 \times 10^6$ & 14696.572$\pm$0.014 & 16290.573$\pm$0.028  \\
$J_4 \times 10^6$ & -586.609$\pm$0.004  & -935.314$\pm$0.037  \\
$J_6 \times 10^6$ & 34.198$\pm$0.009 & 86.340$\pm$0.087 \\
%$C/MR_{eq}^2$ (MOI)$^i$  & 0.264  & 0.22 \\
Rotation Period  & 9h 55m 29.56s &  10h 39m $\pm$10m$^{ii}$\\
Effective Temperature (K) & 124.4$\pm$0.3 & 95.0$\pm$0.4\\
1-bar Temperature (K) & 165$\pm$4 & 135$\pm$5 \\
\noalign{\smallskip}\hline\noalign{\smallskip}
\end{tabular}
\caption{Basic observed parameters of Jupiter and Saturn. The data is taken from  https://ssd.jpl.nasa.gov/ as well as from \cite{RH2018} and references therein. % and NASA website: http://ssd.jpl.nasa.gov/?gravity\_fields\_op. 
The gravitational harmonics for Jupiter and Saturn are taken from \cite{Iess2018} and \cite{Iess2019}, respectively. The gravitational coefficients correspond to the reference equatorial radii of 71,492 km and 60,330 km for Jupiter and Saturn, respectively (http://ssd.jpl.nasa.gov/?gravity\_fields\_op). 
\\
\footnotesize{$^i$these are theoretical values based on interior model calculations. $^{ii}$see \cite{FJ2016} and references therein for discussion on Saturn's rotation rate uncertainty.}
}
%Give details in a table foot note. This is a comment to an entry in the table
}
\end{table}

%%%%%%%%%%%%%%%%%%%%%%%%%%%%%%%
\subsection{Jupiter and Saturn} 
Revealing information on the deep interiors of Jupiter and Saturn must be done by using indirect measurements. Jupiter and Saturn 
%are located at distance that are five and nine times larger than the distance of Earth to the Sun, respectively, and 
are the most massive planets in the Solar System. 
%Jupiter and Saturn are located at distance that are five and nine times larger than the distance of Earth to the Sun, respectively, and therefore, they receive much less heat from the Sun. 
The masses of Jupiter and Saturn are about 318 and 95 times the mass of Earth, respectively. Jupiter's average radius is $\sim$69,911~km, more than ten times Earth's radius. Saturn's  average radius is slightly smaller and measured to be 58,232~km (see Table~2 for details).
%Saturn's mass is nearly 100 times the mass of the Earth and its average radius is smaller than that of Jupiter by a km.  
\par

%Fundamental planetary parameters are the mass and radius, as with this information one can estimate the mean density and therefore the planetary composition. 
The relation between mass and radius provides information about the planet's average density and therefore its bulk composition. % of an object from a measurement of its mass and radius.  
The average densities of Jupiter and Saturn are 1.3262~g/cm$^{3}$ and 0.687~g/cm$^{3}$, respectively. % suggesting that they are mostly composed of H and He. 
The lower average density of Saturn is not a result of a more volatile composition, but due to the smaller effect of compression due to its smaller mass. 
Figure~4 shows a mass-radius (M-R) relation for H-He-rich planets with different compositions and the measured M-R relation of Jupiter and Saturn. 
It can be seen from the figure that both Jupiter and Saturn lie close to the H-He curves, suggesting that they consist of mostly H-He, 
but also have a fraction of heavy elements~\cite{2014arXiv1405.3752G}. 
These heavier elements are typically assumed to be rocks (i.e., silicates and sometimes metals) and/or ices (H$_2$O but also CH$_4$ and NH$_3$). In addition, since Saturn's distance to the pure-H-He curve is larger than that of Jupiter, it is expected to be more enriched with heavy elements. 
Although understanding the internal structure of Jupiter and Saturn relies on the H, and H-He EOS, the heavier elements play key role in the formation and evolution of the planets (see~\cite{2014prpl.conf..643H,HG2018,RH2018} and references therein). 
%Details on that can be found in... bla bla..
%In addition, as discussed above, more complex models with temperature profiles that are non-adiabatic lead to higher fractions of heavies within the planets, which makes them even more important.  
% either in the form of a core, or distributed in the envelope (dotted line). 
%More details can be found in \cite{2014arXiv1405.3752G} and references therein. }

In addition, there are several interesting conclusions about Jupiter and Saturn that can be made simply by plotting their inferred $P$-$T$ profiles  on to of the H phase diagram as shown in Figure~1. First, both planets lie in the regime above solid H, showing that they are indeed fluid planets. % as already suggested by W.~Hubbard in 1968\cite{1968ApJ...152..745H}. 
Second, both planets cross the metallization transition, which is continuous at these temperatures, indicating a smooth transition between 
 molecular and non-to-semiconducting hydrogen in the outer parts of the planets ($H_2$) %which is in the molecular form 
and conducting (or metallic) atomic hydrogen (H) in their deep interiors.  %consisting of metallic (ionic) H. 

The H-He EOS data predetermine the $P$-$T$ profiles of the planets. 
However, the corresponding material data such as compressibility and specific heat have a sensitive impact on the slope of these profiles. 
For instance, the fraction of heavy elements to be mixed with H-He in order to match the measured gravity data and the mass of the core depend on the stiffness of the H-He EOS.   
Therefore, uncertainties in the \textit{ab initio} H-He EOS of $5-10~\%$ as currently valid still have a rather big influence on interior models, and progress in \textit{ab initio} simulations and in high-pressure experiments will directly lead to advanced interior models. 
However, an additional complexity as discussed below is linked to the heat transport mechanism within the planets which is not well determined, and therefore introduces an uncertainly in the planetary temperature profile. 
%Or in other words, the relation between heat conduction/radiation and convection, which is required to determine the internal temperatures.} 
Finally, it is clear that Saturn's {\it P-T} profile covers lower temperatures and pressures (due to its smaller mass which leads to smaller compression). 

%{\color{red} Maybe Ronald? Discuss maybe the error on EOS are of order 5-10 percent if we consider the spread of curves given by DFT with different functionals, and put some consideration about the denser EOS? This will also reinforce the connection between recent findings and planetary science. And Huguniot data to benchmark are still sparse (they are discussed above)}
%As a result, it can be concluded that Saturn's interior consists of a smaller fraction of metallic H in comparison to Jupiter. 
Since the high-P regime of the H EOS is less understood, interior models of Saturn suffer less from the uncertainty in the H EOS.  
This is, however, the opposite when it comes to H-He demixing as indicated by Figure~3. Since Saturn's adiabat crosses the phase separation line is closer its internal structure is more affected by the phase separation of He. This leads to further complexity in modelling Saturn's internal structure, and can also affect its energetics and long-term evolution~\cite{2003Icar..164..228F,LC13,2016ApJ...832..113M,Vazan2016,2016Icar..267..323P,FJ2016}.
\par

Traditionally, internal structure models of Jupiter and Saturn consisted of three distinct layers: a heavy-element core, an inner envelope of metallic hydrogen, and an outer envelope of molecular hydrogen. 
Due to helium separation, the He mass fraction was assumed to be enriched in the inner envelope and depleted in the outer one. The envelopes were also assumed to consist of heavier elements, and their distribution, which was assumed either to be uniformly mixed in both envelopes, or discontinuous  across the two envelopes, was a free parameter in the models. These traditional models assumed that the temperature profile of the planet is given by the adiabatic gradient $\nabla_{ad}$=$\frac{\partial lnT}{\partial lnP}\arrowvert_s$, where $S$ is the entropy. This assumption, however, holds only for a planet which is homogeneous and fully convective. 
It is now clear that both Jupiter and Saturn have more complex interiors, including composition gradients and regions that are stable against convection where heat is transported by conduction or layered-convection \cite{2019ApJ...872..100D, DC19, Vazan18, Vazan2016, LC13, LC12}. 

For Jupiter, accurate gravity data have become available thanks to the Juno mission~\cite{2017Sci...356..821B,Iess2018}. This led to the development of new structure models, with the preferred solutions being ones with dilute cores 
and a discontinuity of the heavy-element enrichment in the envelope, with the inner He-rich envelope consists  
of more heavy elements than the outer, helium-poor envelope~\cite{2017GeoRL..44.4649W,2017A&A...606A.139N,2018Natur.555..227G,DC19}.  
These Jupiter structure models that fit Juno data suggest that giant planets's cores should not be viewed as compact regions consisting mostly of heavy elements but as central regions whose compositions are dominated by heavy elements, which could be gradually distributed or homogeneously mixed. 
A diluted core for Jupiter could extend to a few tens of percents of the planet's total radius, and can also consist of lighter elements, and is actually consistent with the current view of the formation process of gas giant planets~\cite{2017ApJ...840L...4H}. 
It should be noted that the inferred structure of the planets depends on the used EOS. 
%{\bf maybe add that this conclusion depends on the EOS... } 
Also the Cassini mission provided more accurate measurements of Saturn's gravitational field that significantly improved in the last phase of the mission~\cite{Iess2019}.
Saturn structure models that fit Cassini data suggest that Saturn has a relatively massive core of about 15~Earth masses and that its envelope consist of only a few Earth masses of heavy elements~\cite{2019ApJ...879...78M}, with the exact number depending on the assumed rotation period (which is uncertain within a few minutes), and other model parameters~\cite{2013ApJ...767..113H,2019GeoRL..46..616G}. 
These new structure models clearly suggest that the temperature profiles in the planetary deep interior can significantly differ from the adiabatic one. 
This in return can affect the inferred heavy-element mass since hotter interiors lead to higher enrichment for a given density. 
As a result, improvements in  planetary structure models do not only depend on the EOS of the pure materials, but on the more complex interplay of mixtures, phase separations, as well as other physical conditions and processes such as heat transport and convective and mixing efficiencies in the presence of composition gradients.  
Advances in giant planet modeling can also be improved using other data. For example, the splitting of the low order f-modes of Saturn's oscillations  Saturn confirms that the deep interior is non-homogeneous and consists of composition gradients\cite{2014Icar..242..283F}. 
In Figure~3 we present the $P$-$T$ profiles inferred from adiabatic and non-adiabatic models for comparison. 
\par

The much more accurate gravity data has led to the development of more complex structure models that include composition gradients and account for the dynamical contributions. 
Interestingly, the improvements in the quality of the gravity data led to many new questions regarding the planetary interiors, and standard structure models have been challenged. 
It is now clear that the internal structures of Jupiter and Saturn are complex, and are not only dependent on our understanding of the EOS dominating in these objects but also on the heat transport mechanism, the actually internal structure and composition, as well as the thermodynamics and chemistry operating in their deep interiors. 

In addition, the two planets are different from each other, and the challenges in their modelling is not the same - while for Jupiter the largest uncertainty is linked to the H EOS, for Saturn it is the H-He phase diagram and the uncertainty in its rotation period. 
% in modelling each of them is linked to different uncertainties 
% in a mi\citealt{LC12,LC13}  accounted for the possibility of double-diffusive convection in both Jupiter and Saturn interiors caused by heavy-element gradients. It was shown that both 
Figure~6 provides a schematic presentation of the two possible internal structures of Jupiter and Saturn, the traditional structure with three distinct layers, and one with composition gradients (see~\cite{HG2018,2018oeps.book..175H} and references therein for further details). 

%%%%%%%%%%%%%%%%%%%%%%%%%%%%%%%%%%%%%
\subsection{Planetary magnetic fields}
\label{s:magF}
The  fact that Jupiter and Saturn have strong magnetic fields also provides important information on their composition and internal structure. 
In order to possess a magnetic field three criteria should be satisfied: (i) the planet must consist of a region of electrically conductive fluid, (ii) this region should be convective, and (iii) the planet should rotate (at least moderately) rapidly.  
The mechanism for magnetic field generation in planetary interior, also known as dynamo theory, is not well understood and is still being investigated. 
Dynamo generation is related to the physics of MHD and the interplay between fluid dynamics and electro-magnetism and is currently assessed via 
numerical simulations. More information on the connection between high-pressure physics, interior models, and dynamo simulations can be found in {L{\"u}hr} \textit{et al.}~\cite{2018ASSL..448.....L}. 
%As a result, the existence and nature of the magnetic fields provide important observational constraints on their present-day interior structure. 
% in order to generate a dynamo the material must  

Since both Jupiter and Saturn possess dipolar intrinsic magnetic fields, we can conclude that the material in their deep interior is electrically conducting and that at least in part of their interiors, heat is transported through convection.  
Both planets rotate fast enough to generate a dynamo. 
%he existence of intrinsic magnetic field requires large-scale motions in a medium that is electrically conducting (e.g., Robert \& King, 2013). 
For Jupiter and Saturn, the conducting material is thought to be metallic hydrogen. 
The strong connection between planetary interiors and high-pressure physics is therefore clear: from the mean densities of the gas giants can be concluded that they are H-dominated, and since that they possess magnetic fields, one can conclude that H metallises at such high pressures. 
In both Jupiter and Saturn the electrical 
conductivity increases significantly even before full metallization of H is reached~\cite{French2012}. 
The magnitude of the electrical conductivity inside the gas giants, combined with the planetary measured magnetic field strength and luminosity, can be used to estimate the internal Ohmic dissipation and introduce additional constraints for structure and dynamical models (e.g.,~\cite{Liu2008, Cao2017,2014GeoRL..41.5410G,2014Icar..241..148J,2019A&A...629A.125W,2019ApJ...879L..22D}). \\
%\RR{\textbf{We should include papers of Gastine \textit{et al.} and Jones on Jupiter's B field that are based on AIMD data.}} 
%\par
%The formation and evolution Jupiter and Saturn dictate their present-day interior structure and dynamics, which determine the properties of their present-day magnetic field through the magnetohydrodynamic (MHD) dynamo process. 
\indent Jupiter's magnetic field is the strongest in the Solar System (excluding the Sun), and its surface field strength is between 4 and 20~Gauss~\cite{Connerney2018,Moore2018}. 
%Recent Juno observations revealed several surprising factors in the morphology of Jupiter's magnetic field. 
Recently, the Juno spacecraft revealed that Jupiter's magnetic field has an intense isolated magnetic spot near the equator with a negative flux. 
In addition, an intense and relatively narrow band of positive flux near 45 degrees latitude in the northern hemisphere was found together with a rather smooth magnetic field in the southern hemisphere. 
Also the north-south dichotomy in Jupiter's magnetic field structure could be explained by the existence of a diluted core~\cite{Moore2018}. \\
%, which could lead to the generation of the field at higher region and could also leads to the formation of two dynamos within the planet \cite{Moore2018}.  
%This provides a nice link between internal structure models that are based solely on the gravity data and magnetic field measurements. ????}
\indent Saturn's magnetic field which has a surface field strength of 0.2-0.5 Gauss (e.g.,~\cite{2005Sci...307.1266D,2017AGUFM.U22A..02D,2019arXiv191106952C}) is nearly perfectly symmetric with respect to the spin-axis~\cite{2012Icar..221..388C}. 
%\cite{2017AGUFM.U22A..02D,2019arXiv191106952C}
The characteristic of Saturn's magnetic field could be a result of He rain, which could create a stable (against convention) below/above the dynamo. A stable deep interior can also be a result of composition gradients and non-adiabatic interiors. \\
% as discussed above. 
\indent Understanding these processes and their outcomes requires good knowledge of the associated thermodynamics and the feedback on the magnetic field and vice versa.  
% the  However, whether helium rain or composition gradients inside Saturn create substantial stable stratification, and whether this stratification layer is above rather than below the deep dynamo is still being investigated. as discussed above...} 
While the current understanding of the dynamo process is still limited, and as a result, the magnetic fields can only be used to set some bounds on the material properties and heat transports inside the planets. This however, can change in the future.

%%%%%%%%%%%%%%%%%%%%%%%%%%%%%%%
\section{Challenges and Outlook}
The ongoing progress in science leads to the solution of long-lasting open questions, and yet, new questions and challenges arise.  
%Our understanding of the behaviour of elements at planetary conditions and of planetary interiors is still incomplete. Nevertheless, the future seems bright. 
Although our understanding of the giant planets and the behaviour of elements at planetary conditions is still incomplete, we expect significant progress in the near future.
Upcoming experiments and theoretical models are expected to enhance our understanding of phase transitions, mixtures, and immiscibilities. We also foresee  improvements in numerical calculations given the increasing computation power and the development of new numerical techniques. 
In particular, we expect that future experiments will resolve the disagreement on the hydrogen metallization conditions and get consistent results using the various methods. 
In addition, it would be desirable to make experiments of H-He mixtures in order to investigate the demixing of helium in hydrogen.  
Another topic that is expected to blossom in the future is superconductivity. While superconductivity has yet to be found in pure H, the hypothesis of supercontactive H directed the search for superconductivity in H-rich materials~\cite{drozdov2015conventional,liu2017potential}.\\
% where compressed hydride of lanthanum LaH$_{10}$ was recently found to be the warmest superconducting material, with a critical temperature of $\sim 250$~K 
 %(a temperature that can be reached at some points of Earth's surface), 
%at about 170~GPa~\cite{PhysRevLett.122.027001,drozdov2019superconductivity}. 
Finally, in this review we focused on Jupiter and Saturn and have not discussed the ice giants Uranus and Neptune. 
The ice planets are key for understanding planet formation and for the characterisation of intermediate planets around other stars. 
Since these planets are thought to consist of volatiles like water, methane, and ammonia and experimental data focusing on these materials would be valuable. 
In addition, the influence of H-He on the mixtures of these materials and the role of carbon is yet to be determined. 
We also expect progress in our understating of the internal structures 
of Jupiter and Saturn given the ongoing efforts in the processing and the interpretation of recent data from the Juno and Cassini missions, and the development of more comprehensive structure models. 
In addition, upcoming and future space missions will also play a key role in better constraining the interiors of the gas giants. 
The planned ESA JUICE mission will reveal further information on Jupiter, and a potential Saturn probe mission will provide constraints on Saturn's atmospheric composition and the immiscibility of He in H and the process of phase separation.  
Nevertheless, it is now realized that giant planets interiors are much more complex than previously thought. As a result, to better understand giant planets improvements of the H and H-He EOS are required but insufficient. We suggest that future studies should concentrate on mixtures, phase transitions, and their physical properties such as thermal diffusivity, electrical conductivity, and opacity. These properties can then be used to further constrain giant planet formation, evolution, and structure models. 
The link between planetary interiors and high-pressure physics is therefore clear, and we believe that the future holds great promises in this direction. 

%The combination of with giant exoplanet will help us to as prototype... 
%\section{Giant Exoplanet Outlook}
%Finally, the characterisation of giant exoplanets, combined with the knowledge of the Solar System giants, can lead to a more complete understanding of giant planets and the behaviour of materials in planetary conditions. 
%Clearly, we still have not solved all the mysteries related to gaseous planets, and much work is required. However, we expect new observations, exciting discoveries, and theoretical developments that will lead to a leap in understanding the origin, evolution, and interiors of this class of planetary objects.    

%54/101 refs after 2013 or after 2012 but not cited by mcmahon
\subsubsection*{Acknowledgements}
We thank the anonymous referees for valuable comments that helped to improve the manuscript. 
We also acknowledge support from W. Nellis, F.~Soubrian, S.~Sorella, D.~Stevenson, N.~Nettelmann, J.~J.~Fortney, Y.~Miguel, S. M{\"u}ller, C. Valletta, and A.~Cumming. 
RH acknowledges support from SNSF grant 200020\_188460 and thanks the Juno science-team members for inspiring discussions. 
RR acknowledges support by the DFG via the projects FOR 2440 and SPP 1992. 
%David Stevenson, Tristan Guillot and Allona Vazan.... RH acknowledge all the Juno science-team members for inspiring discussions.  

clearpage
\begin{figure}[hbt]
\includegraphics[width=1.0\columnwidth]{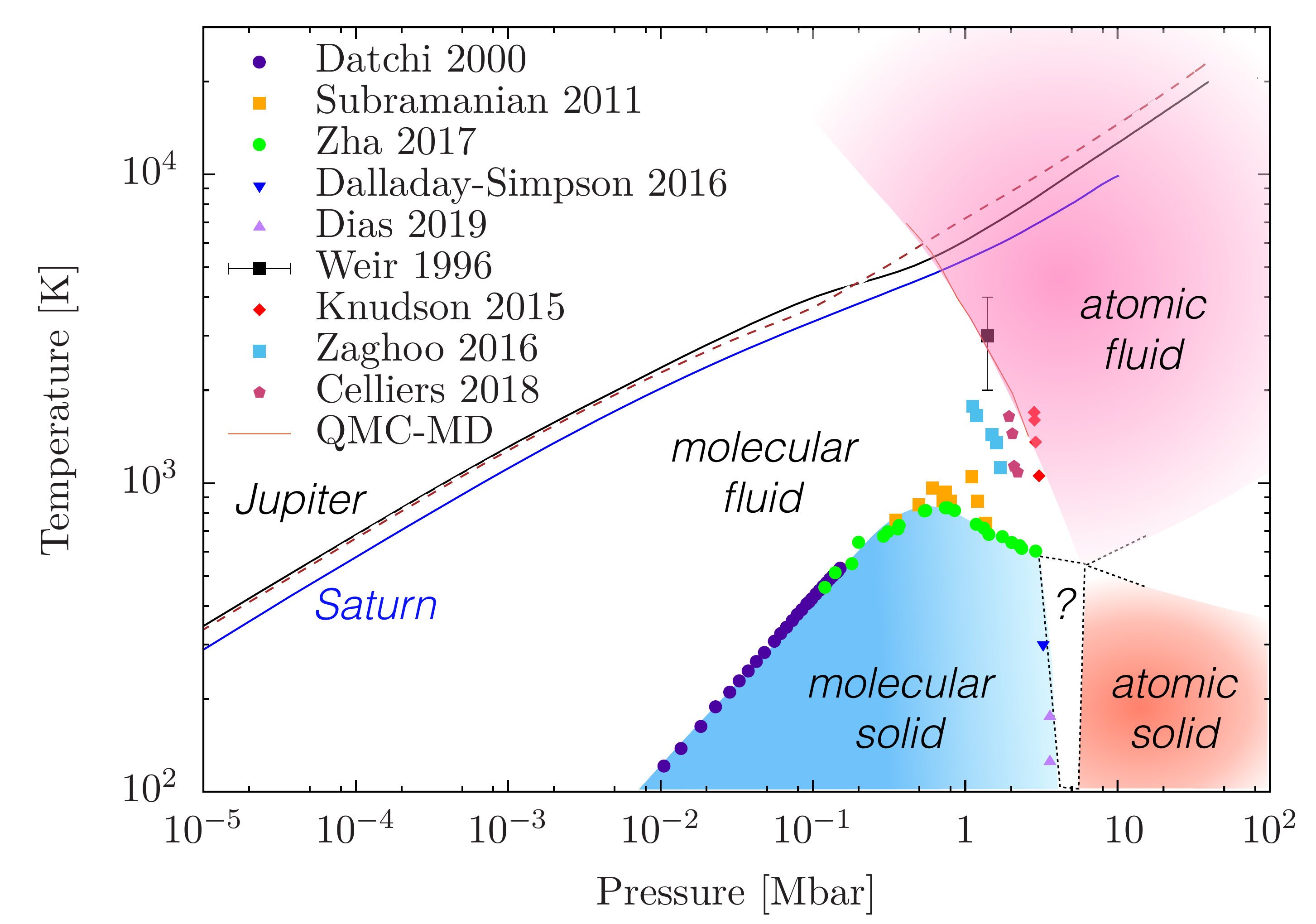}
\vskip -0.25cm
{ \caption{
Phase diagram of hydrogen H (and deuterium D). %thermodynamic conditions of giant planets.  
The solid points represent various experiments performed at the phase boundaries. 
The data coming form Datchi \textit{et al.}~\cite{datchi2000extended}, Subramanian \textit{et al.}~\cite{Subramanian6014} and Zha \textit{et al.}~\cite{PhysRevLett.119.075302} locate the molecular-liquid to molecular-solid ($H_2$) transition. 
%Other phase boundaries separating different $H_2$ solid polymorphs are not shown except 
Also indicated are the measurement of Dalladay-Simpson \textit{et al.}~\cite{dalladay2016evidence} and Dias \textit{et al.}~\cite{dias2019quantum} which represent the highest-pressure phase transitions occurring still in the ($H_2$) solid insulating above 300~GPa.
At finite temperature the transition from the insulating to the metallic solid (red area) remains unknown.
For the liquid phase we show the reported insulator-to-metal transitions form Weir \textit{et al.}~\cite{weir1996metallization}, Knudson \textit{et al.}~\cite{knudson2015direct} ($D$), Zaghoo \textit{et. al.}~\cite{zaghoo2016evidence} ($H$), and Celliers \textit{et al.}~\cite{Celliers677} ($D$). Also plotted are the theoretical prediction of the fluid metallization in the presence of He atoms from Mazzola \textit{et al.}~\cite{mazzola2018} (using electronic QMC with the classical nuclei approximation).
For reference, also shown are typical {\it P-T} for Jupiter (solid black) and Saturn (solid blue), assuming adiabatic interiors~\cite{RH2018}.  
For comparison, we also show a {\it P-T} profile for a non-adiabatic Jupiter where the internal temperatures are higher~\cite{2019ApJ...872..100D}. 
More discussion on structure models is given in section IV.  
 % red one  \cite{RH2018}. }
%{\color{red} ADD the meaning of the new adiabat + ref.}
}}
\end{figure}

\begin{figure}[hbt]
\includegraphics[scale=0.85]{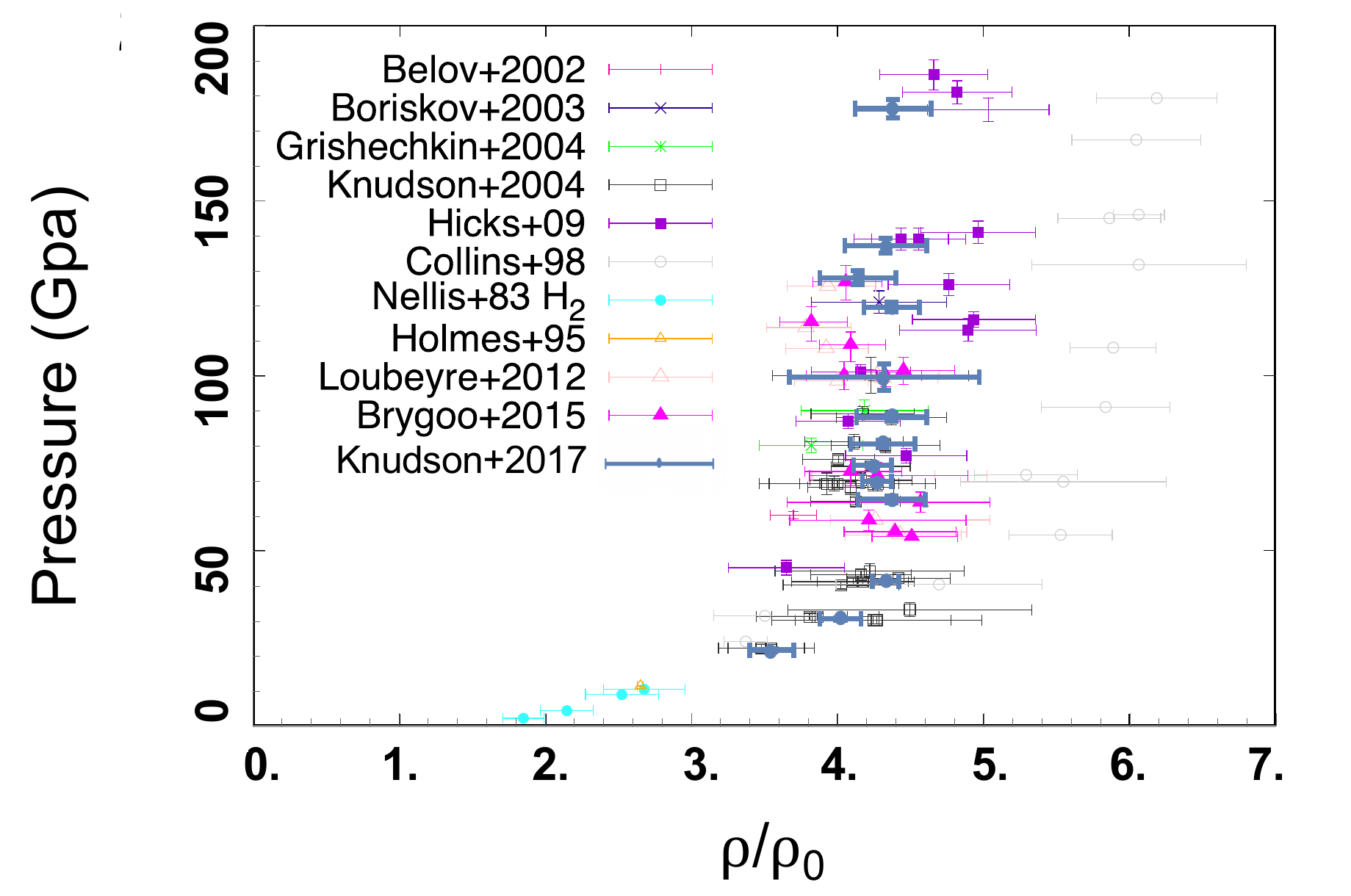}
\vskip -0.4cm
{ \caption{\small Principal Hugoniot data of H. 
%The theoretical calculations correspond to Saumon \textit{et al.}~\cite{saumon1995equation} (SCVH), Becker \textit{et al.}~\cite{becker2014ab} (REOS3b) and  Militzer \& Hubbard \cite{militzer2013ab} (MH13) (which is for a mixture of H-He and therefore in order to get the pure-H He was removed using the SCVH EOS). 
The various experimental results are shown with different point styles for comparison with the references given in the legend. 
The Knudson+2017 data correspond to the weighted inferred by \cite{PhysRevLett.118.035501}. 
The figure is modified from Miguel \textit{et al.}~\cite{Miguel2016}.  
}}
\end{figure}

\begin{figure}
\centering
%{\includegraphics[scale=.27]{H-HE.pdf}}
{\includegraphics[scale=0.6]{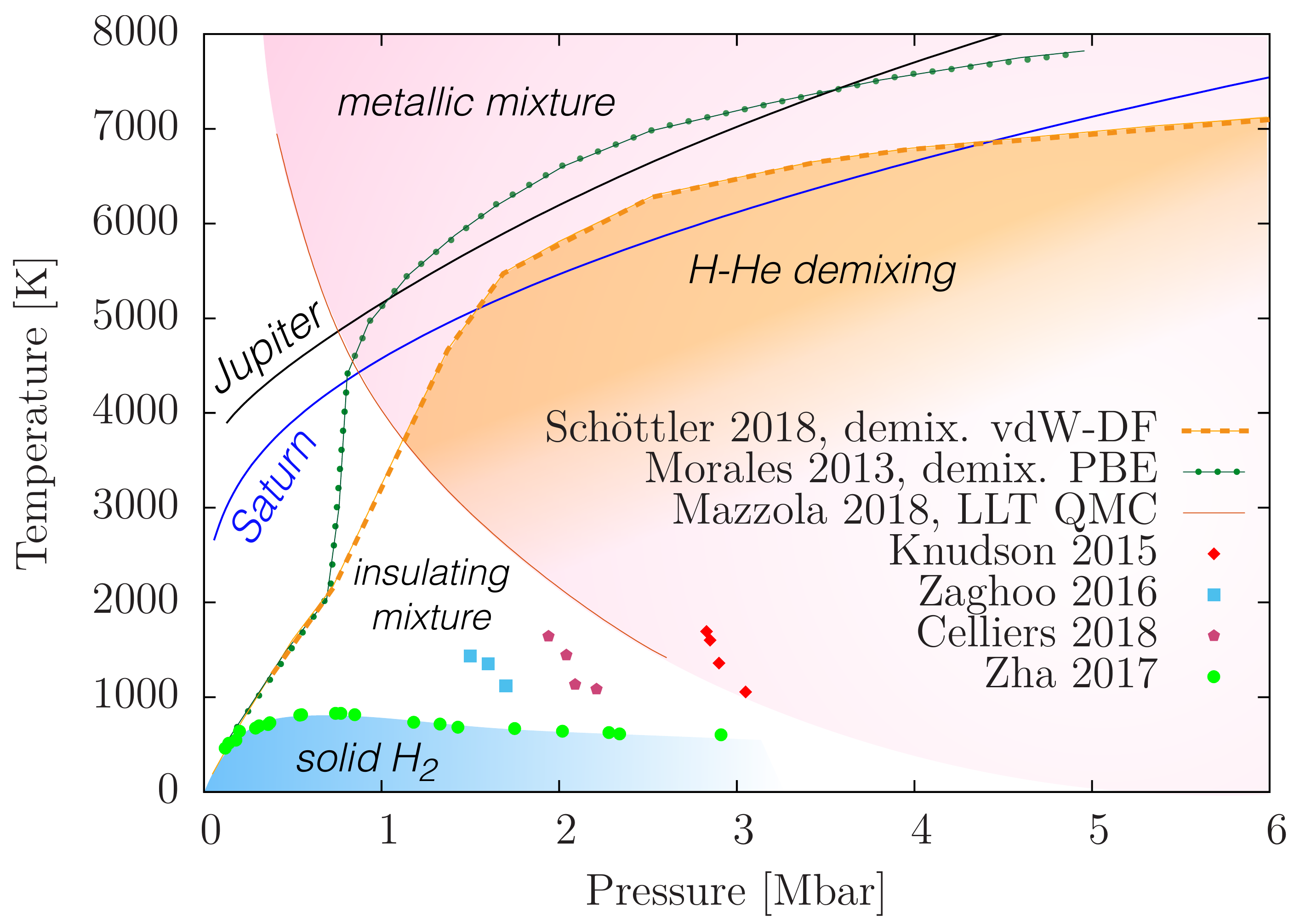}}
%\floatbox[{\capbeside\thisfloatsetup{capbesideposition={left,top},capbesidewidth=4cm}}]{figure}[\FBwidth]
%\vskip -8pt
\vskip -9pt
{\caption{\small 
Phase diagram for a H-He mixture of proto-solar
composition as predicted by numerical calculations together with typical isentropes of Jupiter and Saturn. Orange: demixing region
using the vdW-DF and non-ideal entropy of mixing ~\cite{PhysRevLett.120.115703}, green:
PBE and non-ideal entropy of mixing~\cite{morales2009phase}. The red
line from Mazzola \textit{et al.}~\cite{mazzola2018} separates the insulating and metallic fluid mixture.
Experimental results for the metallization transition in H are
given by coloured symbols: Knudson \textit{et al.}~\cite{knudson2015direct}, Zaghoo \textit{et al.}~\cite{zaghoo2016evidence}, and Celliers \textit{et al.}~\cite{Celliers677}. The H melting line is taken from
Zha \textit{et al.}~\cite{PhysRevLett.119.075302}. 
%For reference also shown are typical isentropes for Jupiter and Saturn. 
%\RR{Phase diagram for hydrogen-helium mixtures of Solar composition as predicted by AIMD. Orange: demixing region within PBE and ideal entropy of mixing~\cite{PhysRevB.84.235109}, magenta: PBE and non-ideal entropy of mixing~\cite{morales2009phase}, red: vdW-DF and ideal entropy of mixing, blue: vdW-DF and non-ideal entropy according to Sch\"ottler \& Redmer~\cite{PhysRevLett.120.115703}. The lower red curves show the melting and the metallization line with its critical point for hydrogen according to Morales \textit{et al.} (2013a). Numerical and experimental results by Schouten \textit{et al.}~(1991) and Loubeyre \textit{et al.}~(1991) are also presented. The black curves show the isentropes of Jupiter (solid) and Saturn (dashed), respectively. % The figure is modified from Sch\"ottler \& Redmer~\cite{PhysRevLett.120.115703}.}
}}
%\vspace{-0.8cm}
\end{figure}

\begin{figure}
\centering
{\includegraphics[scale=0.8]{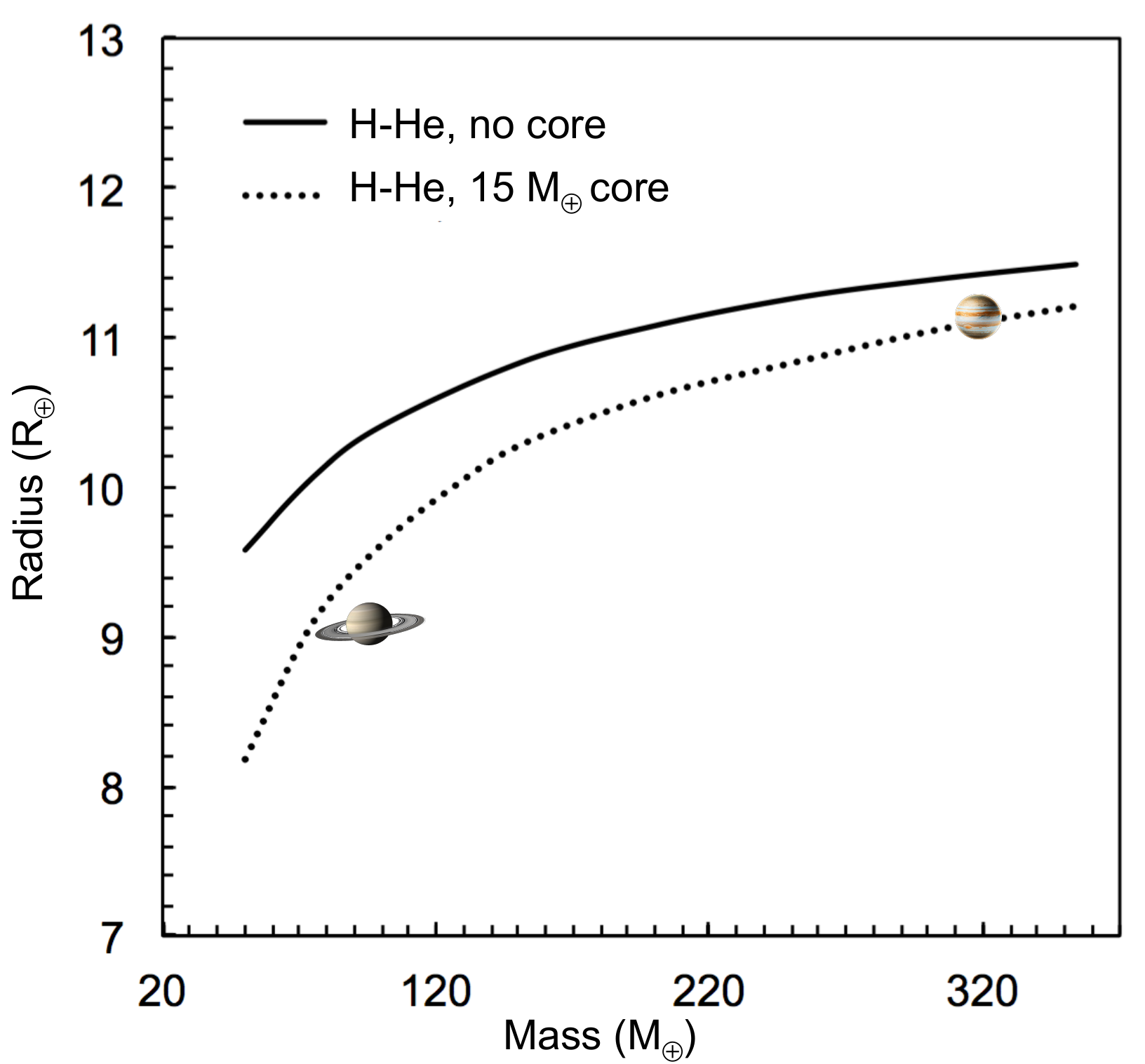}}
%\floatbox[{\capbeside\thisfloatsetup{capbesideposition={left,top},capbesidewidth=4cm}}]{figure}[\FBwidth]
%\vskip -8pt
\vskip -9pt
{ \caption{\small The mass-radius relation of H-He-dominated planets. The solid black curve corresponds to the planets after 5~Ga of evolution composed of pure H-He.
% (solid) and H-He plus a heavy-element core of 15 Earth masses ($_M_{\oplus}$). 
The dotted line shows the effect of a 15~M$_{\oplus}$ heavy-element core on the mass-radius relation. Also shown are Jupiter and Saturn for comparison. 
}}
%The figure is modified from Guillot \& Gautier \cite{2014arXiv1405.3752G}.}}
%\vspace{-0.8cm}
\end{figure}

\begin{figure}
%\floatbox[{\capbeside\thisfloatsetup{capbesideposition={left,top},capbesidewidth=4cm}}]{figure}[\FBwidth]
%\vskip -8pt
%\vspace{-0.8cm}
\centering
%{\includegraphics[scale=.56]{slices_JS.pdf}}
{\includegraphics[scale=.7]{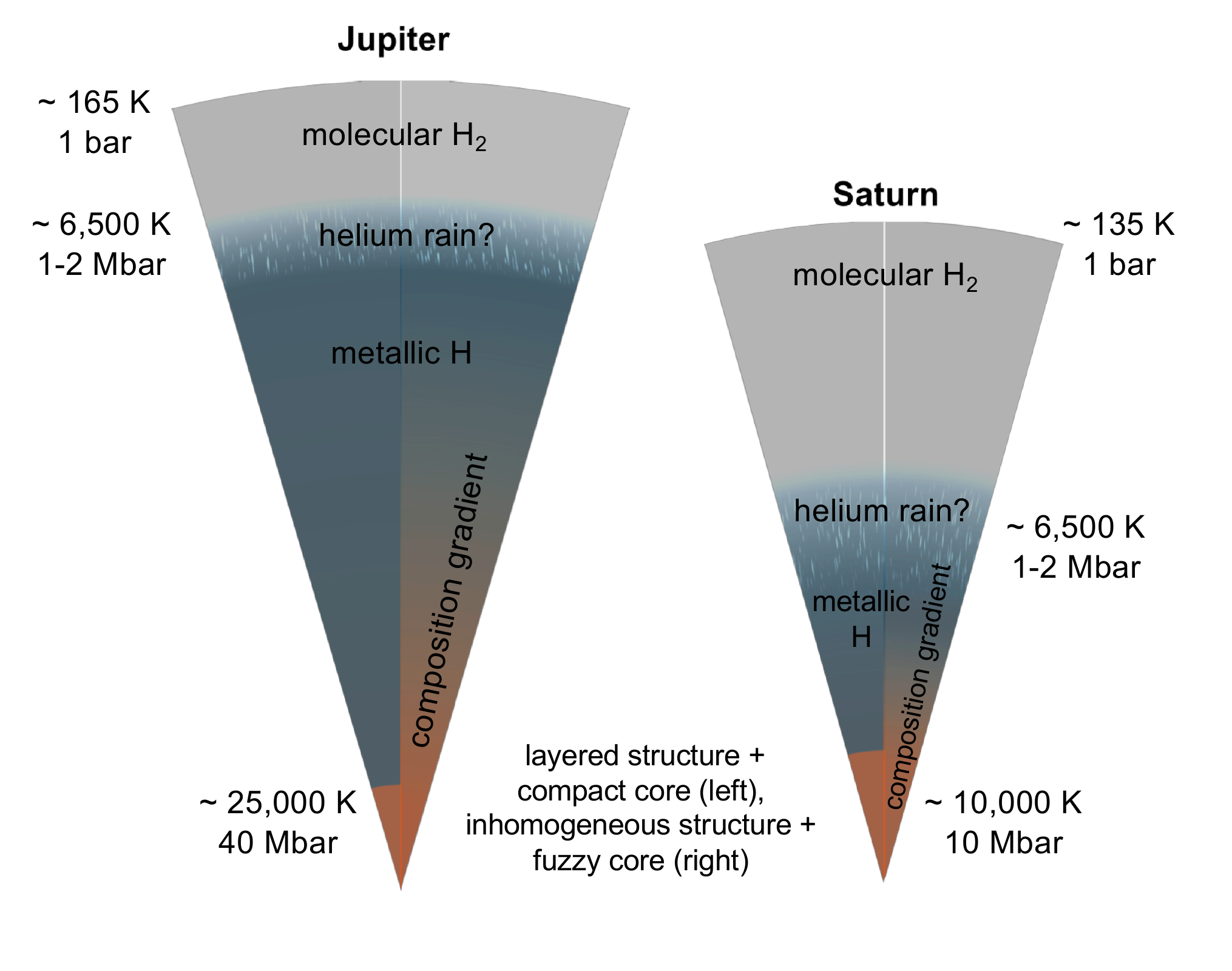}}
{ \caption{\small Sketches of the internal structures of Jupiter and Saturn, modified from Helled, 2018\cite{2018oeps.book..175H}.}}
\end{figure}

\end{document}